%% file: main.tex
\documentclass[12pt]{iopart}

\usepackage[style=authoryear-icomp, %authoryear-ibid,
            maxnames=1,
            minnames=1,
            maxbibnames=3,
            backend=biber,
            url=false,
            uniquelist=false,
            uniquename=false,
            doi=true]{biblatex} % Harvardstyle, acc. PMB

\usepackage{url}
\usepackage[colorlinks=true, 
            citecolor=azure, 
            linkcolor=azure, %figures, anchors etc
            urlcolor=azure]{hyperref} %autoref

\usepackage{orcidlink}
\usepackage{tikz}
\usepackage{pgfplots}
\pgfplotsset{compat=1.18} 
\usetikzlibrary{positioning, calc, decorations.pathreplacing}
\usetikzlibrary{intersections, positioning}
\usetikzlibrary{calc,decorations.pathmorphing}

\usepackage{graphicx, xcolor}
\definecolor{azure}{rgb}{0.0, 0.5, 1.0}

\input{acronyms}

\usepackage[exponent-product = \cdot]{siunitx}
\DeclareSIUnit{\mup}{\text{$\mu_0$}}
\DeclareSIUnit{\mT}{\milli\tesla}
\DeclareSIUnit{\Tpm}{\tesla\per\metre}
\DeclareSIUnit{\mul}{\micro\litre}
\DeclareSIUnit{\degC}{\celsius}    
\DeclareSIUnit{\mgfeml}{\milli\gram\of{Fe}\per\milli\liter}

\usepackage[font={footnotesize}]{caption}

\newcommand{\tu}[1]{\textup{#1}}

\usepackage{nicefrac}
\usepackage{subcaption}

\usepackage{iopams}  
\begin{document}

\title[{Characterization of Resotran for MPI in a Comparison Study}]%kürzer im header
{Characterization of the Clinically Approved MRI
Tracer Resotran for Magnetic Particle Imaging in a Comparison Study}

\author{Fabian Mohn\textsuperscript{1,2}~\orcidlink{0000-0002-9151-9929}, 
Konrad Scheffler\textsuperscript{1,2}~\orcidlink{0000-0002-6842-9204}, 
Justin Ackers\textsuperscript{3}~\orcidlink{0000-0003-1049-0528}, 
Agnes Weimer\textsuperscript{3,4}~\orcidlink{0000-0002-7180-3009},
Franz Wegner\textsuperscript{5}~\orcidlink{0000-0001-5969-3428}, 
Florian Thieben\textsuperscript{1,2}~\orcidlink{0000-0002-2890-5288}, 
Mandy Ahlborg\textsuperscript{3}~\orcidlink{0000-0002-0192-8033}, 
Patrick Vogel\textsuperscript{6}~\orcidlink{0000-0003-0801-5146}, 
Matthias Graeser\textsuperscript{3,7}~\orcidlink{0000-0003-1472-5988}, 
Tobias Knopp\textsuperscript{1,2,3}~\orcidlink{0000-0002-1589-8517}}

\address{\tiny\textsuperscript{1}\scriptsize Institute for Biomedical Imaging, Hamburg University of Technology, Hamburg, Germany}
\address{\tiny\textsuperscript{2}\scriptsize Section for Biomedical Imaging, University Medical Center Hamburg-Eppendorf, Hamburg, Germany}
\address{\tiny\textsuperscript{3}\scriptsize Fraunhofer IMTE, Fraunhofer Research Institution for Individualized and Cell-based Medical Engineering, L\"ubeck, Germany}
\address{\tiny\textsuperscript{4}\scriptsize Institute of Physical Chemistry, University of Hamburg, Germany}
\address{\tiny\textsuperscript{5}\scriptsize Institute for Interventional Radiology, University of L\"ubeck, L\"ubeck, Germany}
\address{\tiny\textsuperscript{6}\scriptsize Department of Experimental Physics 5 (Biophysics), University of W\"urzburg, Germany}
\address{\tiny\textsuperscript{7}\scriptsize Institute of Medical Engineering, University of L\"ubeck, L\"ubeck, Germany}

\ead{\footnotesize \href{mailto:fabian.mohn@tuhh.de}{fabian.mohn@tuhh.de} \hfill February 2024}
\vspace{10pt}
\normalsize

%%%%%%%%%%%%%%%%%%%%%%%%%%%%%%%%%%%%%%%%%%%%%%%%%%%%%%%%%%%%%%%%%%%%%%%%%%%%%%%%%%%%%

\begin{abstract} % PMB style, 250/300 words 
\textit{Objective.} 
The availability of \aclp{MNP} with medical approval for human intervention is fundamental to the clinical translation of \ac{MPI}. In this work, we thoroughly evaluate and compare the magnetic properties of an \ac{MRI} approved tracer to validate its performance for \ac{MPI} in future human trials.
\textit{Approach.}
We analyze whether the recently approved \ac{MRI} tracer Resotran is suitable for \ac{MPI}. In addition, we compare Resotran with the previously approved and extensively studied tracer Resovist, with Ferrotran, which is currently in a clinical phase III study, and with the tailored \ac{MPI} tracer Perimag.
\textit{Main results.}
Initial \acl{MPS} measurements indicate that Resotran exhibits performance characteristics akin to Resovist, but below Perimag. We provide data on four different tracers using \acl{DLS}, \acl{TEM}, \acl{VSM} measurements, \acl{MPS} to derive hysteresis, \acl{PSF}s, and a serial dilution, as well as system matrix based \ac{MPI} measurements on a preclinical scanner (Bruker 25/20 FF), including reconstructed images.
\textit{Significance.}
Numerous approved \aclp{MNP} used as tracers in \ac{MRI} lack the necessary magnetic properties essential for robust signal generation in \ac{MPI}.
The process of obtaining medical approval for dedicated \ac{MPI} tracers optimized for signal performance is an arduous and costly endeavor, often only justifiable for companies with a well-defined clinical business case.
Resotran is an approved tracer that has become available in Europe for \ac{MRI}. In this work, we study the eligibility of Resotran for \ac{MPI} in an effort to pave the way for human \ac{MPI} trials.
\end{abstract}

%%%% Uncomment for keywords
\vspace{2pc}
\normalsize

%%%%%%%%%%%%%%%%%%%%%%%%%%%%%%%%%%%%%%%%%%%%%%%%%%%%%%%%%%%%%%%%%%%%%%%%%%%%%%%%%%%%
%%%%%%%%%%%%%%%%%%%%%%%%%%%%%%%%%%%%%%%%%%%%%%%%%%%%%%%%%%%%%%%%%%%%%%%%%%%%%%%%%%%%
\section{Introduction}
\label{sec:intro}
\acresetall

\Ac{MPI} is an emerging tomographic technique that combines high \ac{MNP} sensitivity with high temporal and spatial resolution~\autocite{gleich_tomographic_2005}. The main principle is the exploitation of the nonlinear magnetization behavior of \acp{MNP} in a periodic magnetic excitation field (drive field). A spatial resolution is achieved by using a magnetic gradient field (selection field) keeping the \acp{MNP} in saturation everywhere except in a small \ac{FFR}.
As a promising tomographic technique without ionizing radiation, \ac{MPI} has high potential in numerous medical applications. Due to its very high temporal resolution, a main focus is cardiovascular and periinterventional imaging~\autocite{weizenecker_three-dimensional_2009,haegele_magnetic_2012,haegele_magnetic_2016,haegele_multi-color_2016,herz_magnetic_2019,wegner_magnetic_2021,bakenecker_magnetic_2018} as well as perfusion imaging~\autocite{ludewig_magnetic_2017,kaul_magnetic_2018,mohn_saline_2023}. Due to the multifaceted properties of \acp{MNP} that can be exploited by \ac{MPI}, further applications are part of extensive research: development of dedicated \ac{MPI} instruments for treatment of vascular stenosis~\autocite{ahlborg_first_2022}, cellular tracking~\autocite{zheng_magnetic_2015,sehl_perspective_2020,remmo_cell_2022}, local magnetic hyperthermia (e.g. tumor imaging and therapy without surgery)~\autocite{chandrasekharan_using_2020,he_simulation_2023} and navigation of magnetic micro-robots for targeted drug delivery and treatment of cerebral aneurysms~\autocite{bui_magnetic_2021,bakenecker_navigation_2021,bui_development_2023}.
The authors refer to reviews for detailed insight on the full functionality of \ac{MPI} as well as the progress made from the first prototype in 2005 to the first commercial preclinical systems, given by~\cite{knopp_magnetic_2017}. Further outlines over current research and applications can be found in~\cite{billings_magnetic_2021,yang_applications_2022,neumann_recent_2022}.

Besides upscaling \ac{MPI} hardware to human-sized scanners~\autocite{sattel_setup_2015,rahmer_remote_2018,graeser_human-sized_2019,mason_design_2017,vogel_impi_2023} and addressing safety concerns~\autocite{thieben_system_2023,schmale_human_2013,saritas_magnetostimulation_2013}, the availability of suitable \acp{MNP} with medical approval is crucial for a clinical translation of \ac{MPI}. The development of medical \acp{MNP} is primarily driven for the application in \ac{MRI}. Unfortunately, most \acp{MNP} developed for \ac{MRI} do not have the specific magnetic properties that are needed to generate a strong signal in \ac{MPI}. 
If nanoparticles are too small, the thermal energy dominates the magnetic energy, inducing a rather linear magnetization behavior. Thus, they are not suited for \ac{MPI}, where signal generation and spatial encoding is based upon a nonlinear magnetization. On the other hand, if particles are too large, they block the Neél relaxation process due to strong magnetic anisotropies. This reduces their ability to follow the magnetic field at excitation frequencies between 10 kHz to 150 kHz~\autocite{deissler_dependence_2014, tay_relaxation_2017}.
An important \ac{MRI} \ac{MNP} that has magnetic properties suitable for \ac{MPI} is ferucarbotran, namely Resovist, formerly with medical approval in Germany (Bayer Schering Pharma, Berlin, Germany) and still approved in Japan (I'rom Pharmaceuticals, Tokyo, Japan). However, due to a wide particle size distribution with the majority of particles being smaller than \SI{15}{\nm}, only a small fraction of the total iron mass contributes to a useful \ac{MPI} signal. First dedicated \acp{MNP}, tailored to enhance the \ac{MPI} specific performance, were published by~\cite{ferguson_optimization_2009}. Later, a monodisperse iron core \ac{MNP} coated with polyethylene glycol~\autocite{ferguson_magnetic_2015}, evolving into the formerly commercially available \acp{MNP} LS-008 (LodeSpin Labs, Seattle, USA) was developed. In 2013, dextran coated multicore magnetic iron oxide nanoparticles were presented by~\cite{eberbeck_multicore_2013}, commercially available as the preclinical \acp{MNP} Perimag and Synomag (micromod Partikeltechnologie, Rostock, Germany). 
Moreover, \acp{MNP} can also undergo a system-specific optimization, i.e., to match a particular type of excitation: the formation of particle chains has improved the nonlinear response in 1D excitation~\autocite{tay_superferromagnetic_2021}.  
A comparison of commercial \acp{MNP} with respect to their \ac{MPI} performance is given by~\cite{ludtke-buzug_comparison_2013} and~\cite{ludwig_characterization_2013} and more recently by~\cite{irfan_development_2021}. A recent overview of the development of \ac{MPI} tailored \acp{MNP} is given by~\cite{harvell-smith_magnetic_2022}.

The research on \ac{MPI} tailored tracers increased in the last years~\autocite{antonelli_development_2020,liu_long_2021,moor_particle_2022,thieben_development_2023}, however, none of these tracers has reached a level of development that would warrant the costs of a medical approval and consequently their use in clinical \ac{MPI} remains distant. Such an approval requires a well-defined business case and a long-term market to justify the multi-annual process and investment in a new approval. 
Fortunately, Resotran (b.e.imaging GmbH, Baden-Baden, Germany; medical approval granted in Oct. 2022 under reg. no. 7002837.00.00 in Germany), containing ferucarbotran \acp{MNP}, has recently received approval in certain countries, including Germany and Sweden. Additionally, there is a phase III clinical trial underway for ferumoxtran \acp{MNP} called Ferrotran, consisting of \ac{USPIONs}. Both \acp{MNP} are officially authorized for \ac{MRI} liver imaging and initial measurements showed similar \ac{MPI} performance~\autocite{hartung_resotran_2023,scheffler_mpi_2023}.
General concerns regarding toxicity of \acp{MNP} in long-term metabolism remain~\autocite{sun_magnetic_2008,billings_magnetic_2021,rubia-rodriguez_whither_2021}, although the incidence of adverse events for ferucarbotran (Resovist) is low with \SI{7.1}{\%} ~\autocite{wang_superparamagnetic_2011}.

The purpose of this paper is to provide a comprehensive characterization of the \acp{MNP} Resotran and Ferrotran with a focus on their applicability to \ac{MPI}. Comparisons will be made with the extensively studied \ac{MRI} \acp{MNP} Resovist as well as with the \ac{MPI} tailored \acp{MNP} Perimag. We chose Perimag because of its established position and its appearance in a wide range of publications and open datasets~\autocite{knopp_openmpidata_2020}.
We address the characterization of the four \acp{MNP} by \ac{TEM}, \ac{DLS}, \ac{VSM} and \ac{MPS} measurements. In addition, 2D \ac{MPI} reconstructions for two different phantoms are compared at the system matrix level and in the image domain for future applications in \ac{MPI}. We present and discuss the results of applying these methods to Resotran, Ferrotran, Resovist, and Perimag.

%%%%%%%%%%%%%%%%%%%%%%%%%%%%%%%%%%%%%%%%%%%%%%%%%%%%%%%%%%%%%%%%%%%%%%%%%%%%%%%%%%%%
%%%%%%%%%%%%%%%%%%%%%%%%%%%%%%%%%%%%%%%%%%%%%%%%%%%%%%%%%%%%%%%%%%%%%%%%%%%%%%%%%%%%
\section{Materials and Methods}
\label{sec:methods}

For a comprehensive characterization of the four \acp{MNP} Perimag, Resotran, Resovist and Ferrotran regarding their suitability in \ac{MPI}, we analyze shape parameters, magnetic properties, system matrix performance and image reconstructions.

First, the hydrodynamic diameter can be determined using \ac{DLS} and the core diameter of the magnetite can be determined using \ac{TEM}. The latter provides a detailed visualization of the inner iron core in a sub-nanometer resolution and thus of the relationship between the iron structure and performance in \ac{MPI}.
Second, regarding the magnetic properties, we determine the static magnetization characteristic by \ac{VSM} and the dynamic particle response to a drive field by \ac{MPS}. The \ac{VSM} data are used to observe the \acp{MNP} in the saturation region as well as their nonlinear slope through the origin according to the Langevin model. \ac{MPS} measurements show the particle spectrum and can reveal relaxation induced hysteresis as a function of excitation amplitude. We also measure a dilution series and different offset-field combinations to plot two types of \acp{PSF} that can be used to estimate image resolution.
Third, prior to reconstruction, the \ac{SNR} and system matrix patterns are analyzed to estimate the performance and compare Resotran to Resovist in the frequency domain. The fineness of the frequency pattern indicates the expected resolution of the reconstructed image.
Finally, the \acp{MNP} are evaluated in a typical \ac{MPI} imaging scenario to demonstrate suitability and resolution for medical imaging, using a commercial imaging system (Bruker MPI 25/20 FF). Two different phantoms are measured and we also perform cross reconstructions using the Resotran system matrix to reconstruct all other tracers to assess compatibility.

In the following we introduce each of these methods in detail and describe the performed experiments and their implications.

%%%%%%%%%%%%%%%%%%%%%%%%%%%%%%%%%%%%%%%%%%%%%%%%%%%%%%%%%%%%%%%%%%%%%%%%%%%%%%%%%%%
\subsection{Magnetic Nanoparticle Material}
\label{sec:methods:trcr}

The \acp{MNP} are measured at a concentration of $\SI{8.5}{\mgfeml}\approx\SI{152}{\milli\mol/\liter}$, a threshold that is chosen to avoid concentration dependent behavior~\autocite{lowa_concentration_2016}. For Perimag, we use stock dispersion with this concentration (LOT 045211). Both Resotran (LOT F1901) and Resovist (LOT 20F01) are supplied with \SI{28}{\mgfeml} and are therefore diluted with distilled water. Ferrotran (LOT PRX19L02) is shipped as freeze-dried powder and has a concentration of \SI{20}{\mgfeml} once dispersed in water, which we dilute to the same level of \SI{8.5}{\mgfeml}.
All \acp{MNP} are made from iron oxide and coated with a dextran shell. More specifically, Resovist and Resotran are made from Ferucarbotran and Ferrotran is made from Ferumoxtran-10 Lyophilisate and additionally coated with sodium citrate.

%%%%%%%%%%%%%%%%%%%%%%%%%%%%%%%%%%%%%%%%%%%%%%%%%%%%%%%%%%%%%%%%%%%%%%%%%%%%%%%%%%%
\subsection{Dynamic Light Scattering}
\label{sec:methods:DLS}

The hydrodynamic diameter of the aqueous iron oxide nanoparticle dispersion is measured using \ac{DLS} on a Zetasizer Pro-Blue (Malvern Panalytical Ltd., Malvern, United Kingdom) device at a laser wavelength of \SI{633}{\nm}. The sample is diluted 1:100 with Milli-Q (Merck Group, Darmstadt, Germany) water and measured in a plastic cuvette at an optical path length of \SI{1}{\cm}. Each measurement is recorded over three cycles (3 averages) of \SI{30}{\s} each and an intensity weighted mean hydrodynamic diameter of the particle ensemble (z-average) is calculated with the respective \ac{PDI}. The z-average is based on the method of cumulants~\autocite{koppel_analysis_1972}, where the monochromatic light source is scattered by the \acp{MNP} in suspensions and the light intensity of the interference pattern is evaluated for a logarithmic normal size distribution~\autocite{thomas_determination_1987}. The light scattering is caused by the particle ensemble surface and the results include the dextran shell, therefore a size distribution of the hydrodynamic diameter is shown, not the magnetite core.
The data is analyzed using the ZS XPLORER software version 3.2.0.84 (Malvern Panalytical Ltd., Malvern, United Kingdom).

%%%%%%%%%%%%%%%%%%%%%%%%%%%%%%%%%%%%%%%%%%%%%%%%%%%%%%%%%%%%%%%%%%%%%%%%%%%%%%%%%%%
\subsection{Transmission Electron Microscopy}
\label{sec:methods:TEM}

\ac{TEM} measurements are performed with a JEOL JEM-1011 (JEOL Ltd., Tokyo, Japan) at \SI{100}{\kV} equipped with two spherical aberration correction devices (CETCOR and CESCOR by CEOS GmbH, Heidelberg, Germany) and a Gatan 4K UltraScan 1000 (Gatan Inc., Pleasanton, USA) camera.
For the preparation, \SI{10}{\micro\liter} of the diluted nanoparticle dispersion are placed on a carbon-coated \ac{TEM} copper slide with a \SI{400}{\mm} mesh. The excess solvent is removed with a filter paper and the \ac{TEM} grid is air-dried. The recorded images achieve a \SI{2e5}{}-fold magnification at \SI{100}{\kV}. For a quantitative analysis, the size of $250$ individual particles is measured using the software ImageJ (NIH, Bethesda, USA) and plotted in a histogram to visualize the size-distribution, following the guidelines of \cite{iso_13322-1_static_2014} for counting. Our evaluation only accounts for the short-axis diameter~\autocite{verleysen_evaluation_2019,pyrz_particle_2008} of individual particles and we do not count any particle clusters or chains~\autocite{bresch_counting_2022}.

%%%%%%%%%%%%%%%%%%%%%%%%%%%%%%%%%%%%%%%%%%%%%%%%%%%%%%%%%%%%%%%%%%%%%%%%%%%%%%%%%%%%%%
\subsection{Vibrating Sample Magnetometer}
\label{sec:methods:vsm}
The magnetization of the liquid samples in response to static magnetic fields are characterized using a \acl{VSM} (Lakeshore 8607 VSM, Westerville, USA). A quantity of \SI{20}{\ul} is filled into the sample holder and covered with oil, resulting in an almost spherical sample shape. A sweep of the external magnetic field in the range of $\pm\SI{2}{\tesla}$ (step size \SI{20}{\mT}) and in the range of $\pm\SI{30}{\mT}$ (step size \SI{0.5}{\mT}) is performed. The signal is averaged for \SI{1}{\s} at each step. Results are given in the domain of the magnetic moment, calibrated by the \ac{VSM}~\autocite{foner_versatile_1959} and divided by 2 to match the iron mass of the \ac{MPS} samples of \SI{85}{\micro\g\of{Fe}} (\SI{10}{\micro\liter}).

%%%%%%%%%%%%%%%%%%%%%%%%%%%%%%%%%%%%%%%%%%%%%%%%%%%%%%%%%%%%%%%%%%%%%%%%%%%%%%%%%%%%
\subsection{Magnetic Particle Spectroscopy}
\label{sec:methods:MPS} 

We use an arbitrary waveform \ac{MPS} to measure different \SI{10}{\micro\l} samples of \acp{MNP} exposed to a combination of a static and a dynamic magnetic field~\autocite{mohn_system_2022}. These fields are homogeneous inside the measurement chamber and consist of two quantities, a sinusoidal drive-field $B_\textup{drive}$ at \SI{26.042}{\kHz} and a static offset field $B_\textup{offset}$ for saturation. In this case, both fields are oriented in the same direction. A set of static offsets in the range of $\pm\SI{30}{\mT}$ (step size \SI{0.5}{\mT}) is measured for different drive-field amplitudes in the range of $\SIrange{4}{20}{\mT}$ (step size \SI{2}{\mT}). All measurements are averaged over $45$ drive-field periods (\SI{1.73}{\ms}) to reduce noise at low drive-field values.
The receive bandwidth of the \ac{MPS} device is \SI{7.8125}{\MHz}, using a stack of two \acl{RP} and the open source software stack composed of RedPitayaDAQServer~\autocite{hackelberg_flexible_2022} and MPIMeasurements.jl~\autocite{hackelberg_mpimeasurementsjl_2023}. The system is calibrated using a transfer function measured with a small calibration coil~\autocite{thieben_receive_2023}. By calibrating the entire receive chain, we can express the particle response in terms of the net magnetic moment $m$ and thus obtain device-independent measurements that are particle specific. 
The hysteresis curve is obtained by plotting $m$ against the actual drive-field $B_\textup{drive}$, using the calibrated reference channel in \si{\mT} of the device.

%---------------------------------------
\subsubsection{Point Spread Function.}
\label{sec:methods:MPS:PSF}

Two types of \acp{PSF} are calculated to visualize tracer differences using the \ac{MPS} data. A narrow and steep \ac{PSF} is generally indicative for high resolution \ac{MPI}~\autocite{croft_relaxation_2012}, while relaxation effects cause asymmetries and broadening of the \ac{PSF}.
The dynamic \ac{PSF} is based on a straight forward approach by plotting one half-cycle of $\frac{\textup{d} m}{\textup{d} t}$ against the excitation $B_\textup{drive}$ (positive half-cycle only). Consequently, the \ac{PSF} approaches a zero-crossing at the maximum amplitude of $B_\textup{drive}$.
The calculation of the $x$-space \ac{PSF} is typically based on partial \acp{FOV} and a DC-recovery step~\autocite{goodwill_x-space_2012,lu_linearity_2013}, which becomes obsolete when the fundamental is not filtered, i.e. when using a gradiometric arbitrary waveform \ac{MPS}, as validated by~\cite{tay_high-throughput_2016}. To this end, we plot the value of $\frac{\textup{d} m}{\textup{d} t}$ at the maximum field gradient of $B_\textup{drive}$ against each offset step value. The data is then normalized to facilitate the comparison of the \ac{FWHM} of the $x$-space \ac{PSF}.

%---------------------------------------
\subsubsection{Serial Dilution.}
\label{sec:methods:MPS:VDR}

To investigate the linearity between the particle magnetization and the total amount of iron in a sample, we perform a dilution series with an \ac{MPS} with 1D sinusoidal excitation with \SI{20}{\milli\tesla} at \SI{26.042}{\kHz}. Each measurement is performed using \num{10} background frames and \num{10} foreground frames, using a transfer function correction and a sample of \SI{10}{\micro\liter} of each \ac{MNP}. Starting with \SI{8.5}{\mgfeml} the concentration is halved seven times, dispersed with the same amount of distilled water, leading to a set of 8 measurements per tracer with $8.5\,\cdot\,\left(\nicefrac{1}{2}\right)^i \si{\mgfeml} \text{ for } i=0, \dots, 7$. Despite working with highest precision, potential inaccuracies while pipetting increase with a diminishing total iron amount. We evaluate the absolute values of the third harmonic of the measured magnetization response in the frequency domain to compare the results of the 4 different \acp{MNP}.

%%%%%%%%%%%%%%%%%%%%%%%%%%%%%%%%%%%%%%%%%%%%%%%%%%%%%%%%%%%%%%%%%%%%%%%%%%%%%%%%%%%%
\subsection{Magnetic Particle Imaging}
\label{sec:methods:mpi}

\ac{MPI} is performed using the preclinical Bruker MPI system 25/20 FF. We use a 2D Lissajous excitation in $xy$-direction with an amplitude of $\SI{12}{\mT}$ at a frequency of \SI{24509}{\kHz}\,/\,\SI{26041}{\kHz} in $x$-/$y$-direction and a selection field gradient of $(-1,-1,2)$~\si{\Tpm} generating a \ac{FOV} of \qtyproduct[product-units = power]{24 x 24}{\milli\meter}. All measurements are taken with a dedicated 3D receive coil with an open bore of \SI{72}{\mm}, based on the gradiometric approach, a custom built \acl{LNA}, and corrected with a measured transfer function~\cite{graeser_towards_2017}.
The 2D system matrices are measured using a delta-sample of \qtyproduct[product-units = power]{1 x 1 x 5}{\milli\meter} filled with $\SI{4}{\micro\liter}$ of each tracer diluted to a common iron concentration of $\SI{8.5}{\milli\gram_{Fe}\per\milli\liter}$ on $29 \times 29 \times 1$ equidistant grid positions covering \qtyproduct[product-units = power]{29 x 29 x 5}{\milli\meter}. A quantitative comparison of the different \acp{MNP} on system matrix level is done by considering the \ac{SNR} profiles and characteristics~\autocite{franke_system_2016} as well as the \ac{SSIM}~\autocite{wang_image_2004} over all frequency components. Furthermore, a qualitative comparison is given on two selected frequency components with high (\SI{100.98}{\kilo\hertz}) and low (\SI{105.57}{\kilo\hertz}) \ac{SNR}.

The \ac{MPI} reconstructions are performed on two different phantoms, each measured with \num{500} averages (\SI{10}{\s} measurement time). The first phantom consists of three \qtyproduct[product-units = power]{1 x 1 x 1}{\milli\meter} square samples, filled with \SI{0.8}{\micro\liter} of the tracer at \SI{8.5}{\mgfeml}, placed in the corners of an equilateral triangle with an edge length of \SI{9.24}{\milli\meter}. If the \acp{MNP} are \ac{MPI} suitable, the individual dots should be easily separable. The second phantom is more complex and consists of a spiral with two full windings. The round vessel has a diameter of \SI{2.5}{\mm} and a minimal distance to the next winding of \SI{2.8}{\mm}, also filled with a concentration of \SI{8.5}{\mgfeml}. Although the total iron amount is much higher than in the three-dot phantom, a complete resolution of the spiral is expected to be more challenging than for the three-dot phantom. Image reconstruction is performed with the iterative Kaczmarz method and a careful selection of frequency components and regularization strength.

%%%%%%%%%%%%%%%%%%%%%%%%%%%%%%%%%%%%%%%%%%%%%%%%%%%%%%%%%%%%%%%%%%%%%%%%%%%%%%%%%%%
%%%%%%%%%%%%%%%%%%%%%%%%%%%%%%%%%%%%%%%%%%%%%%%%%%%%%%%%%%%%%%%%%%%%%%%%%%%%%%%%%%%
\section{Results}
\label{sec:res}

All measurements of \autoref{sec:methods} were performed identically for the four considered \acp{MNP}. With the exception of the dilution series and the DLS experiments, all \acp{MNP} were prepared at identical concentrations of \SI{8.5}{\mgfeml} (\SI{152}{\milli\mol\per\liter}).

%%%%%%%%%%%%%%%%%%%%%%%%%%%%%%%%%%%%%%%%%%%%%%%%%%%%%%%%%%%%%%%%%%%%%%%%%%%%%%%%%%%
\subsection{Dynamic Light Scattering}
\label{sec:res:DLS}

\begin{figure}[t!]
    \centering
    \includegraphics[width=0.95\linewidth]{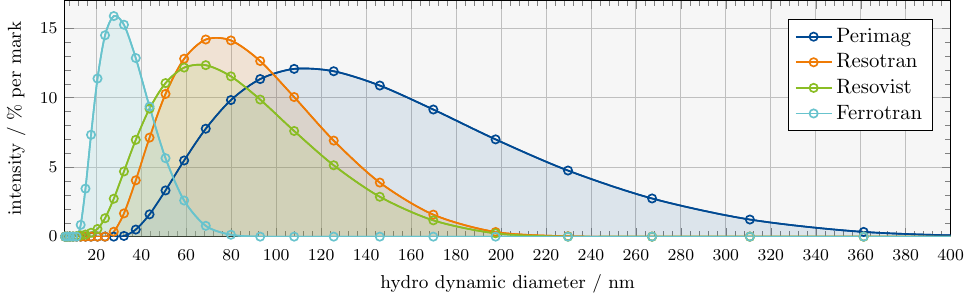}
    \caption{\textbf{Intensity weighted log normal size distribution by DLS.} A sample with a 1:100 dilution of each tracer was measured using DLS to determine the hydrodynamic particle diameter, that includes the dextran shell coating (laser wavelength of \SI{633}{\nm}, \SI{30}{\s} measurement time, $3$ averages).}
    \label{fig:DLS_results}
\end{figure}

\begin{figure}[t!]
    \centering
    \includegraphics[width=1.0\linewidth]{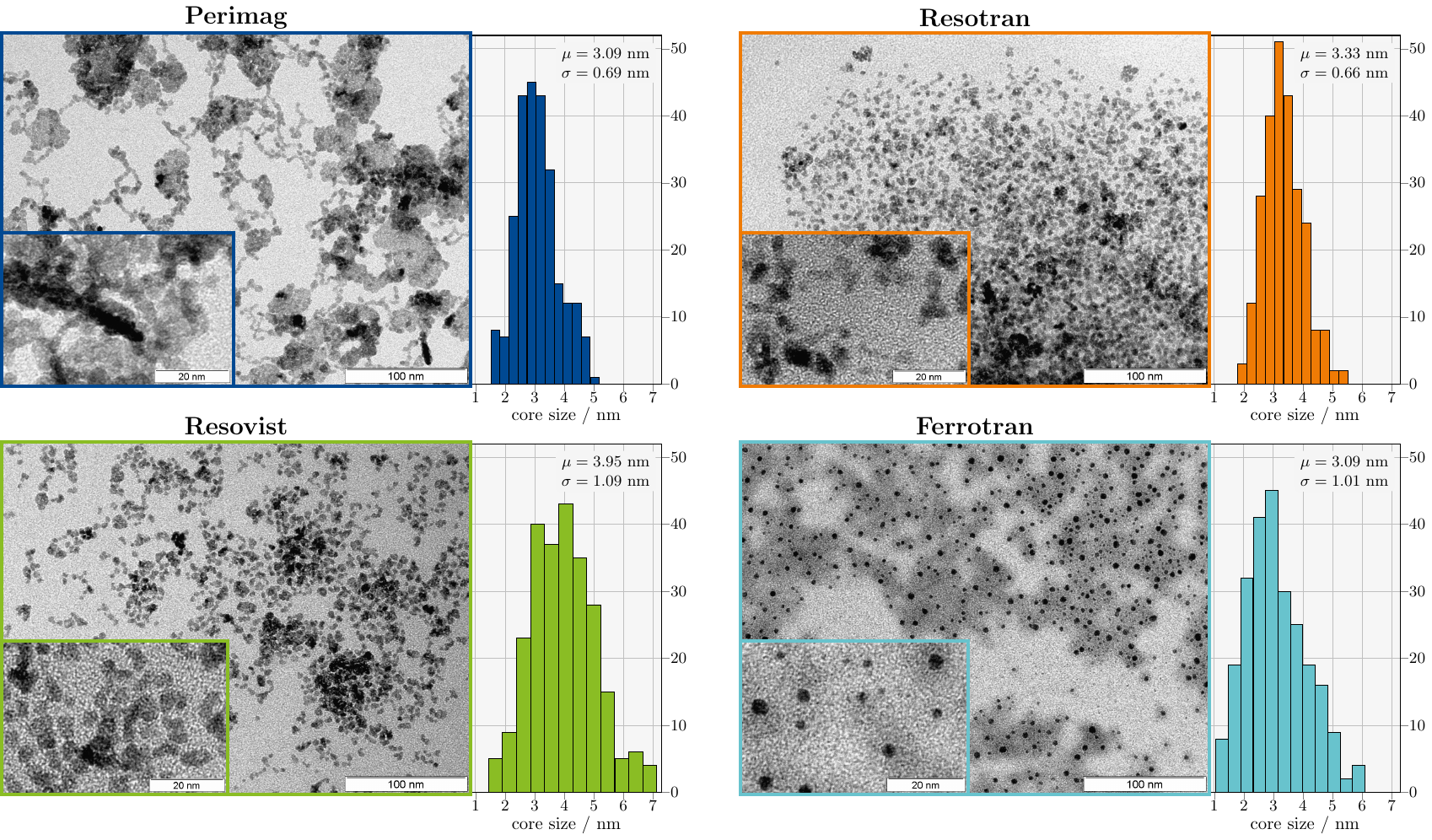}
    \caption{\textbf{TEM images.} Samples of each tracer as visualized by TEM for two different zoom levels at \SI{20}{\nm} and \SI{100}{\nm} reference scale, showing the iron-oxide core. Each histogram lists the mean $\mu$ and the standard deviation $\sigma$ of the core short-axis diameter for a count of $n=250$ particles categorized into $12$ individual size bins. Note, that Perimag, Resotran and Resovist form clusters and chains, which were not counted, whereas Ferrotran accounts for the smallest and individual particles (\ac{USPIO}).}
    \label{fig:TEM_results}
\end{figure}

Results of the \ac{DLS} measurements are shown in \autoref{fig:DLS_results}. Light intensity is given in percent for each size bin (round marks) with respect to all measured bins of the log normal size distribution.
Perimag exhibits the largest hydrodynamic sizes with a peak value at \SI{114}{\nm} (z-average \SI{102.5}{\nm}, \ac{PDI} 0.1853), followed by Resotran with a narrower distribution and a peak at \SI{74}{\nm} (z-average \SI{66.32}{\nm}, \ac{PDI} 0.1806). Resovist is roughly comparable to Resotran, with a peak value at \SI{65}{\nm} (z-average \SI{55.97}{\nm}, \ac{PDI} 0.2007). Ferrotran is a \ac{USPIO} and has a hydrodynamic diameter of around \SI{28}{\nm} (z-average \SI{27.82}{\nm}, \ac{PDI} 0.09) and the narrowest distribution of all tracers.

%%%%%%%%%%%%%%%%%%%%%%%%%%%%%%%%%%%%%%%%%%%%%%%%%%%%%%%%%%%%%%%%%%%%%%%%%%%%%%%%%%%
\subsection{Transmission Emission Microscopy}
\label{sec:res:TEM}

In \autoref{fig:TEM_results}, \ac{TEM} images and a histogram of individual particles for a count of $n=250$ short-axis core diameter measurements are shown for each tracer. The mean $\mu$ and the standard deviation $\sigma$ are given in the top right corners. \ac{TEM} images provide indications of shape, structure, size and uniformity of the nanoparticles. 
Note that the counting rule applied significantly influences the classification~\autocite{bresch_counting_2022}, but we mostly observe spherical individual particles without strong elongation and do not classify particle clusters.

Perimag and Ferrotran exhibit the smallest mean core diameter, followed by Resotran and Resovist in increasing order. As \ac{TEM} images show the magnetite cores, clusters and particle-chains are visible as well as overlapping particles.
Especially for Perimag, such clusters and chains are visible in the pictures and the cores tend to form large clusters in the range of \SIrange{20}{50}{\nm}. Visual inspection of Resotran and Resovist indicates a similar structure and size of both tracers. Ferrotran shows isolated cores, separated by their ligand hull, which reduces magnetic interaction in between particles. Seemingly, no clusters are formed and particles do not overlap, which agrees with \ac{DLS} results for this \ac{USPIO}.

%%%%%%%%%%%%%%%%%%%%%%%%%%%%%%%%%%%%%%%%%%%%%%%%%%%%%%%%%%%%%%%%%%%%%%%%%%%%%%%%%%%
\subsection{Vibrating Sample Magnetometer}
\label{sec:res:VSM}

\begin{figure}[b!]
    \centering
    \includegraphics[width=0.95\linewidth]{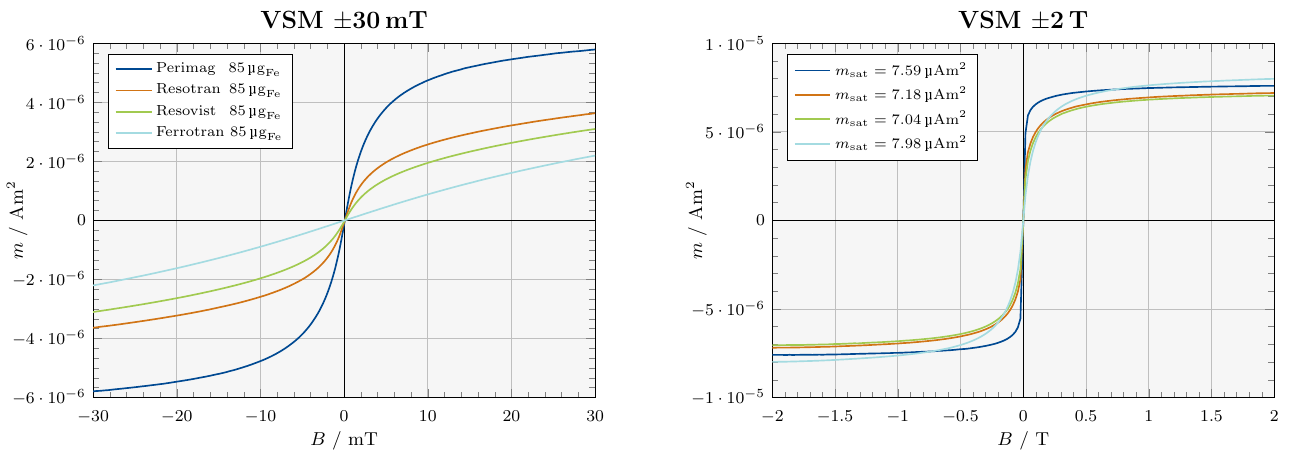}
    \caption{\textbf{VSM results.} The magnetic moment is plotted against the external field for two different ranges with identical samples. The steeper the slope of the net magnetic moment $m$ at the origin, the stronger is the nonlinear response which is the useful signal in \ac{MPI}.}
    \label{fig:VSM_results}
\end{figure}

The results of the VSM measurements are shown in \autoref{fig:VSM_results}. All particle samples show the expected superparamagnetic behavior with sigmoidal magnetization curves and no detectable hysteresis. In the $\pm\SI{2}{\tesla}$ range plot, the magnetization curves of Resovist and Resotran are very similar, reaching almost the same saturation magnetization at about \SI{83.6}{\ampere\meter^2\per\kg_{Fe}}. The saturation magnetization of Perimag is around \SI{89.3}{\ampere\meter^2\per\kg_{Fe}}. 
%which is consistent with the value given in the product datasheet \textcolor{blue}{[todo cite]}. 
Ferrotran has the highest saturation magnetization of the investigated particles (\SI{93.88}{\ampere\meter^2\per\kg_{Fe}}), but has a much lower initial slope with an almost linear curve in the MPI relevant range $\pm\SI{30}{\mT}$, when the origin of the left plot in \autoref{fig:VSM_results} is considered. In contrast, Perimag shows the strongest nonlinearity with an initial slope around \SI{1.23}{\micro\ampere\meter^2\per\mT}, whereas Resotran exhibits \SI{0.67}{\micro\ampere\meter^2\per\mT} and Resovist is lower with \SI{0.45}{\micro\ampere\meter^2\per\mT} (evaluated at $\pm\SI{1}{\mT}$).

%%%%%%%%%%%%%%%%%%%%%%%%%%%%%%%%%%%%%%%%%%%%%%%%%%%%%%%%%%%%%%%%%%%%%%%%%%%%%%%%%%%
\subsection{Magnetic Particle Spectroscopy}
\label{sec:res:MPS}

\begin{figure}[b!]
    \centering
    \includegraphics[width=1.0\linewidth]{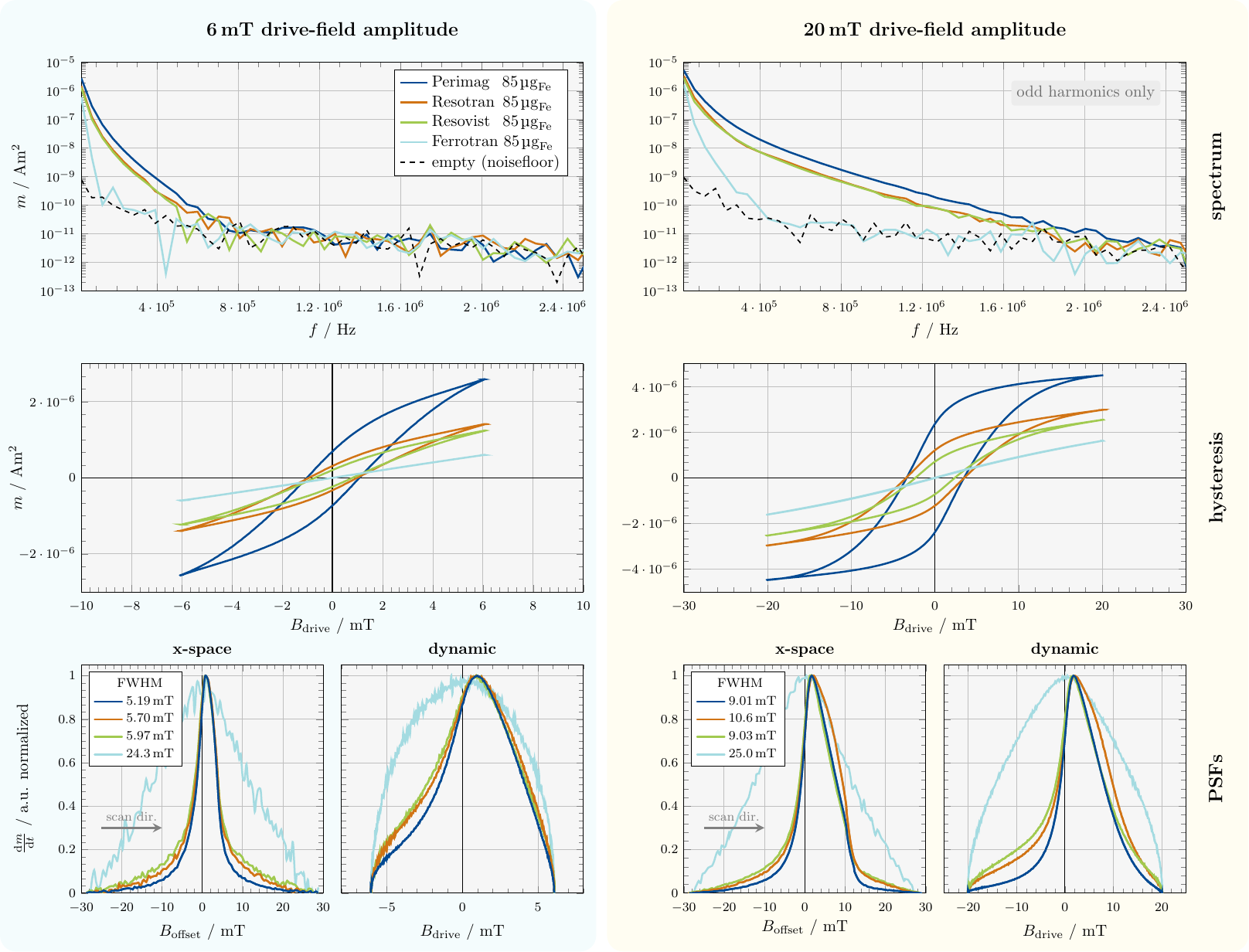}
    \caption{\textbf{Arbitrary waveform MPS results.} Spectrum, hysteresis and \acp{PSF} from top to bottom for \SI{6}{\mT} and \SI{20}{\mT} sinusoidal drive-field excitation. The frequency spectrum only contains odd harmonics of the fundamental $f_1=\SI{26.042}{\kHz}$. The \ac{FWHM} of the $x$-space PSF is given in the legend and the gray arrow indicates the scan direction (positive half-wave).}
    \label{fig:pMPS_results}
\end{figure}

Four types of plots are generated in \autoref{fig:pMPS_results}, each containing measurements of the four \acp{MNP} under investigation. On the top, the spectrum for \SI{6}{\mT} and \SI{20}{\mT} excitation amplitude is shown, because these amplitudes refer to a realistic range for human-sized \ac{MPI}~\autocite{ozaslan_pns_2022,thieben_system_2023}. We only plotted the odd harmonics, because they contain the majority of the information on the nonlinear magnetization of \acp{MNP} in a homogeneous sinusoidal excitation field (without offset fields). The spectrum of Resotran and Resovist are very similar for both \SI{6}{\mT} and \SI{20}{\mT}, the amplitude of Resotran being slightly higher in the range of \SIrange{5}{15}{\percent}. Compared to Resotran, Perimag has a $1.5$ to $2.0$-fold higher signal amplitude at low harmonic indices, increasing to $2.5$ at higher indices above \SI{400}{\kHz}. Ferrotran has an overall low signal amplitude, as already indicated by the linear slope in \autoref{fig:VSM_results}. Even at \SI{20}{\mT} excitation, useful signal is only detectable below \SI{300}{\kHz}, indicating insufficient \ac{MPI} signal.

In the middle row, the hysteresis curve is plotted, which shows a residual magnetization for all tracers, except for Ferrotran, which does not seem to undergo a measurable, relaxation-induced hysteresis.
On the bottom the different normalized \acp{PSF} are plotted. We state the \ac{FWHM} for the $x$-space \ac{PSF} to facilitate comparison. At \SI{6}{\mT}, all tracers except Ferrotran indicate very similar magnetic properties, however, the difference in terms of \ac{FWHM} between both \ac{PSF} types are large. This effect reduces at \SI{20}{\mT} with a trend for the $x$-space \ac{PSF} to broaden and the dynamic \ac{PSF} to narrow.
Both \acp{PSF} indicate an inferior \ac{MPI} image quality for Ferrotran. The noisy shape is caused by the normalization, which maps all peak amplitudes to one. Their original relation of maximum signal can be deduced from the saturation region of the hysteresis curve above.

\begin{figure}[h!]
    \centering
    \includegraphics[width=1.0\linewidth]{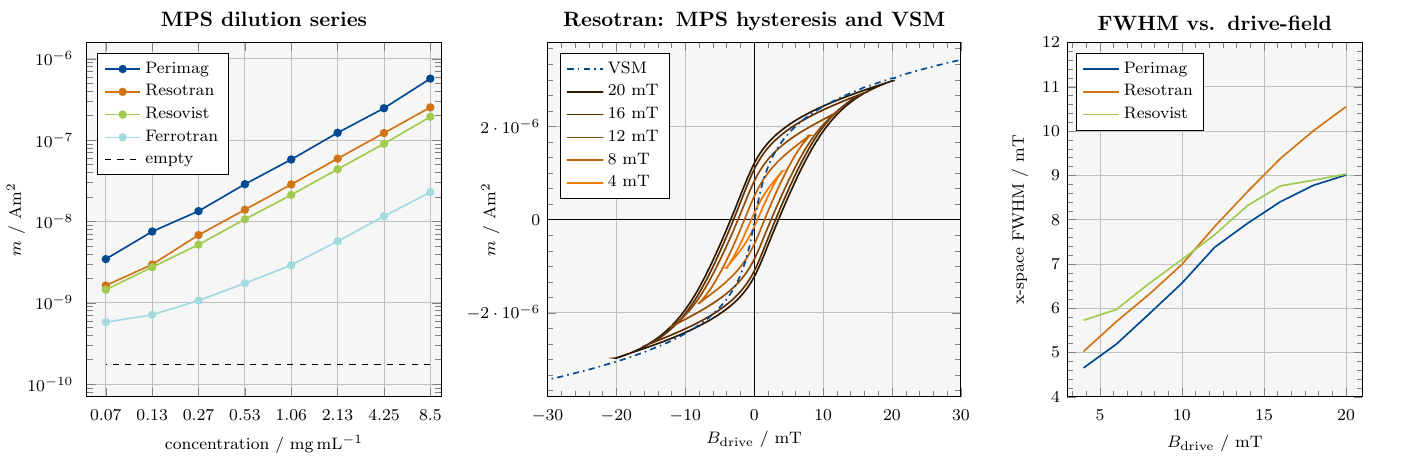}
    \caption{\textbf{MPS data analysis.} On the left, a dilution series of all tracers is shown. The absolute value of the third harmonic of exponentially decreasing concentrations between $8.5\cdot\left(\nicefrac{1}{2}\right)^0$\,\si{\milli\gram\per\milli\liter} and $8.5\cdot\left(\nicefrac{1}{2}\right)^7$\,\si{\milli\gram\per\milli\liter} is plotted (\SI{20}{\milli\tesla}, \SI{26.042}{\kHz}). The center plot focuses on Resotran: the hysteresis for 5 different excitation amplitudes is plotted (bright to dark for increasing amplitude) with an overlay of the VSM curve (dash-dot). On the right, the \ac{FWHM} is plotted against the drive-field amplitude. Ferrotran was omitted to retain a detailed scale.}
    \label{fig:MPS_analysis}
\end{figure}

In a deeper \ac{MPS} analysis, we refer to three different types of plots in \autoref{fig:MPS_analysis}. 
On the left side, the results of the \ac{MPS} dilution series are shown. We observe a linear behavior in the absolute magnitude of the third harmonic over all concentrations down to  $8.5\cdot\left(\nicefrac{1}{2}\right)^7\approx0.066~\si{\milli\gram\per\milli\liter}$ for Perimag, Resotran and Resovist. While Perimag produces the highest signal, the results indicate that Resotran and Resovist are relatively comparable in signal strength, with Resotran's signal being slightly higher. Ferrotran gives a much weaker \ac{MPS} signal, starting with a linear result for higher concentrations, but loosing linearity by the 4th dilution step towards lower concentrations. At the lowest two concentrations, the response is barely higher than the background signal.

In the center of \autoref{fig:MPS_analysis}, the hysteresis curve for Resotran is plotted in a range of \SIrange{4}{20}{\mT} with \SI{4}{\mT} steps for the excitation fields. In addition, the \ac{VSM} curve of Resotran is overlaid, as obtained from \autoref{fig:VSM_results}. The hysteresis broadens with increasing amplitude and the turning point in saturation (maximum/minimum of $B_\tu{drive}$) seems to approach the \ac{VSM} line at high amplitudes.

The $x$-space \ac{PSF} was evaluated at varying excitation amplitudes between \SIrange{4}{20}{\mT} in \SI{2}{\mT} intervals, as shown on the right-hand side of \autoref{fig:MPS_analysis}. The resulting plot displays the \ac{FWHM} of the \acp{PSF} against the drive-field amplitude and provides detailed insights into its tendency to increase with amplitude, which was previously observed in \autoref{fig:pMPS_results}. Specifically, the \ac{FWHM} seems to increase linearly with amplitude. However, the linear increase is only applicable to a small region, and both Perimag and Resovist demonstrate a decline above \SI{16}{\mT}. On the other hand, Resotran maintains a consistent linear pattern throughout our measurement range.

%%%%%%%%%%%%%%%%%%%%%%%%%%%%%%%%%%%%%%%%%%%%%%%%%%%%%%%%%%%%%%%%%%%%%%%%%%%%%%%%%%%
\subsection{Magnetic Particle Imaging}
\label{sec:res:mpi}

\begin{figure}[t!]
    \centering
    \includegraphics[width=0.9\linewidth]{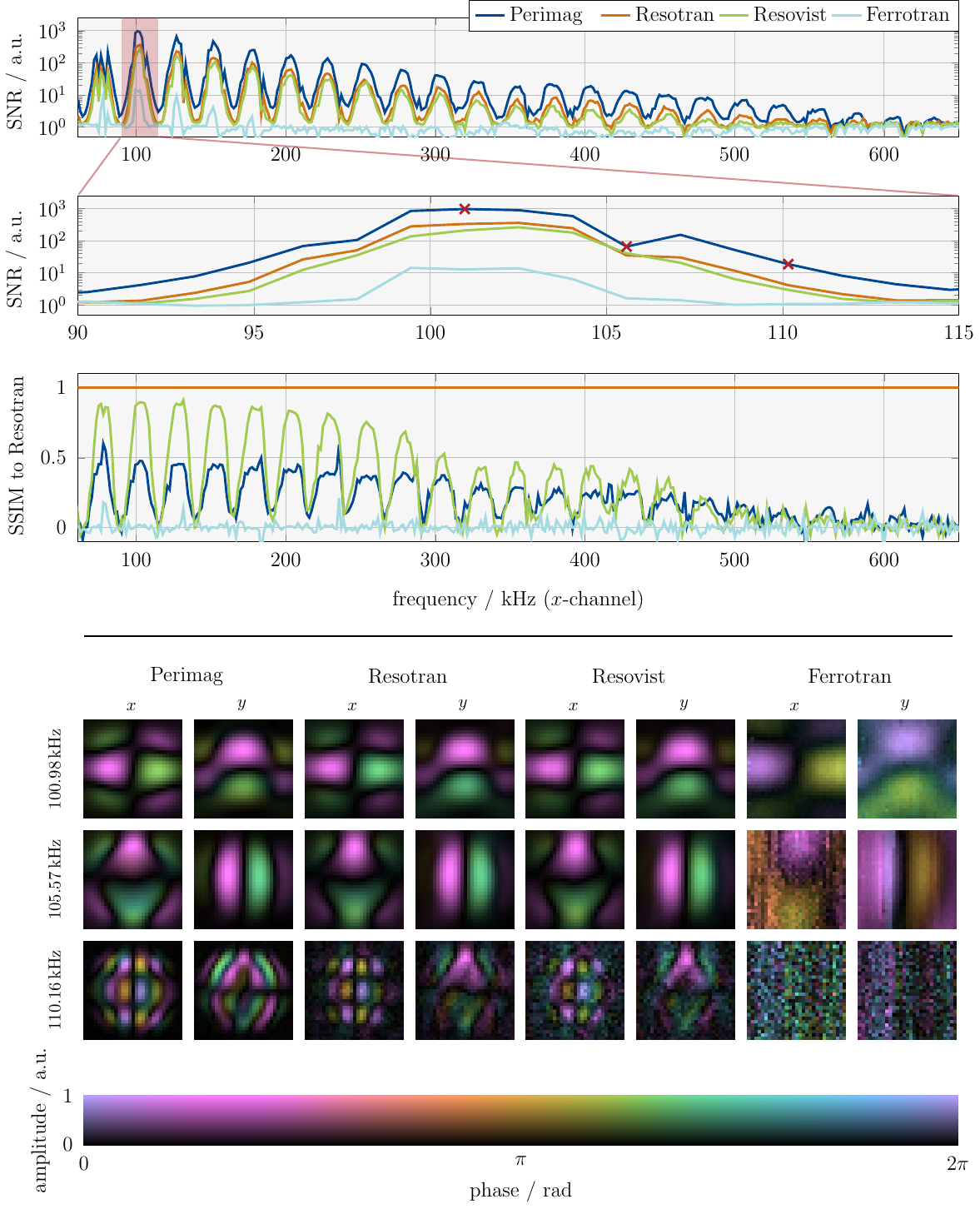}
    \caption{\textbf{MPI system matrices.} On the top, the \ac{SNR} of the $x$-channel of all \acp{MNP} is shown, as measured with the preclinical system Bruker MPI 20/25 FF with $xy$-excitation. The relation of signal amplitudes and visible harmonics is similar to the \ac{MPS} measurements. The second row, shows a zoom on the fourth harmonic (\SI{101}{\kHz}) including its side bands. The third row displays the \ac{SSIM} of Perimag, Resovist and Ferrotran with respect to Resotran for the full spectrum. Below, the complex color-coded system matrix pattern are shown for three frequencies (marked by red crosses above) across all tracers.}
    \label{fig:MPI_SM}
\end{figure}

For a quantitative comparison of the system matrices, we consider the \ac{SNR} profile of the $x$-channel as well as an \ac{SSIM} comparison with Resotran over all frequencies of the $x$-channel, shown in the upper part of \autoref{fig:MPI_SM}. As expected, Perimag achieves the highest \ac{SNR} over the entire frequency band. Especially for higher harmonics above \SI{350}{\kilo\hertz} the measured signal outperforms the other tracers. Ferrotran clearly shows the lowest \ac{SNR} profile. Only the first 4 harmonics reach \ac{SNR} levels suitable for \ac{MPI}. In contrast, Resotran and Resovist achieve \ac{SNR} profiles suitable for good MPI measurements with useful frequency components up to \SI{500}{\kHz}. Higher harmonics show good \ac{SNR} levels above $10$ up to \SI{350}{\kilo\hertz}. Overall, Resotran and Resovist show very similar \ac{SNR} profiles. This is supported by the \ac{SSIM} comparison in the 3rd row of \autoref{fig:MPI_SM}. Moreover, the \ac{SSIM} indicates, that the Resotran and Resovist system matrix patterns are similar in structure. Especially around the harmonics with a \ac{SNR} above 10, the \ac{SSIM} is high and the system matrix patterns of Resotran and Resovist are very similar. 

A qualitative comparison of the system matrices on three selected frequency components with high \ac{SNR} (\SI{100.98}{\kilo\hertz}), medium \ac{SNR} (\SI{105.57}{\kilo\hertz}) and low \ac{SNR} (\SI{110.16}{\kilo\hertz}) is displayed in the bottom part of \autoref{fig:MPI_SM}. The close resemblance of the system matrix components of Resotran and Resovist can be seen in phase and amplitude. Moreover, the wave patterns show the same structure when compared to Perimag. Visible differences compared to Perimag can be seen in the component with medium \ac{SNR}, especially in the outer corners and in the component with low \ac{SNR}, where Resotran and especially Resovist have significantly more noise. At the highest frequency we also see Resotran and Resovist differing in quality. The system matrix components of Ferrotran are clearly different and the wave patterns are not represented correctly even for the component with the high \ac{SNR} value. Phase and amplitude are also different and noise is clearly dominant.

\begin{figure}[t!]
    \centering
    \includegraphics[width=1.0\linewidth]{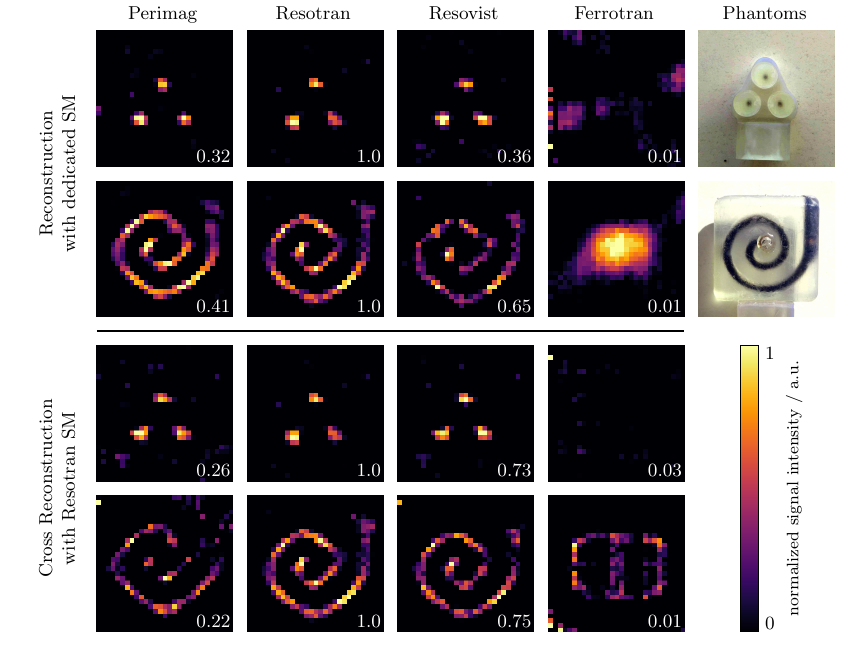}
    \caption{\textbf{Comparison of MPI reconstructions.} Results of the four different \acp{MNP} in two different phantoms are shown, each with an iron concentration of \SI{8.5}{\mgfeml}. The signal intensity is normalized for each image individually. The MPI tailored tracer Perimag provides the best image quality. Resotran and Resovist are similar, but Ferrotran has a very weak particle response. The \ac{SSIM} to the reconstruction unsing Resotran (row-wise) is superposed in the lower right corner.}
    \label{fig:MPI_recos}
\end{figure}

Lastly, we consider the reconstruction results of two phantoms, a 3-dot phantom and a spiral phantom, given in \autoref{fig:MPI_recos}. In the upper part, reconstructions with the dedicated system matrices are shown. In the lower part, cross-reconstructions with the Resotran system matrix are shown for all other tracers. The \ac{SSIM} to the reconstruction result of Resotran is superposed in the lower right corner for all reconstructions. The images indicate, that both phantoms can be reconstructed successfully by using Perimag, Resotran and Resovist. Here, Perimag visibly achieves the best image quality, followed by Resotran. Ferrotran is not able to resolve the phantoms at all. Cross-reconstructions using the Resotran system matrix are possible for Perimag and Resovist on both phantoms. While the result for Perimag gets worse, the reconstruction result for Resovist improves when using the Resotran system matrix. This is supported by the increasing \ac{SSIM} to the reconstruction of the Resotran spiral ($0.75$) in comparison to the reconstruction of the spiral using the Resovist system matrix ($0.65$). One reason may be that there are more disruptions and noise in the Resovist system matrix, as it can be seen in the lowest system matrix component in \autoref{fig:MPI_SM}.

%%%%%%%%%%%%%%%%%%%%%%%%%%%%%%%%%%%%%%%%%%%%%%%%%%%%%%%%%%%%%%%%%%%%%%%%%%%%%%%%%%%
%%%%%%%%%%%%%%%%%%%%%%%%%%%%%%%%%%%%%%%%%%%%%%%%%%%%%%%%%%%%%%%%%%%%%%%%%%%%%%%%%%%
\section{Discussion}
\label{sec:discuss}

This work evaluates the four particles Perimag, Resotran, Resovist, and Ferrotran by classifying their sizes, magnetic properties, and imaging performance. We have shown that Resotran and Resovist are similar in composition and performance, with acceptable imaging results, and that Ferrotran is unsuitable for \ac{MPI}.

The purpose of this study is threefold: first, to establish a relationship of the well-known and thoroughly studied \acp{MNP} Resovist to the newly \ac{MRI} approved \ac{MNP} Resotran, both ferucarbotran, to indicate similar properties and \ac{MPI} performance. Secondly, this work makes a contribution on the way to an official approval of Resotran for human applications in \ac{MPI} for vascular imaging based on its similarity to Resovist. Toxicological risks of \acp{MNP} on the metabolism remain~\autocite{winer_use_2012,singh_potential_2010,chen_inhibitory_2010}, however, the long-term application of ferucarbotran in human \ac{MRI}~\autocite{reimer_application_1995,amemiya_dynamic_2009} since its introduction in 2003~\autocite{reimer_ferucarbotran_2003} has so far not raised any major concerns~\autocite{wang_superparamagnetic_2011}.
Third, we complement our study with two more \acp{MNP}, one \ac{MPI} tailored tracer called Perimag to indicate possible future increases in performance and dosage and another called Ferrotran, \ac{USPIO}s currently in a Phase III study, but not suitable for \ac{MPI} due to their linear magnetization behavior caused by their low magnetic energy and their small core size.
We chose Perimag, because it is well studied in the literature~\autocite{eberbeck_multicore_2013,ludtke-buzug_comparison_2013} over a period of 10 years, however, other tracers such as Synomag~\autocite{gavilan_colloidal_2017,vogel_synomag_2021}, VivoTrax (Magnetic Insight Inc., Alameda, United States), PrecisionMRX (Imagion Biosystems Ltd, Melbourne, Australia)~\autocite{tay_relaxation_2017}, LS-008~\autocite{vogel_micro-traveling_2019}, or magnetosomes~\autocite{makela_magnetic_2022,thieben_development_2023} have also shown significant potential and even superior magnetic properties for \ac{MPI} compared to Perimag~\autocite{irfan_development_2021,yeo_characterizing_2022}.

The size discrepancy of \ac{DLS} and \ac{TEM} measurements shows how the hydrodynamic diameter reveals clusters and chains: although Perimag and Ferrotran have similar individual mean core diameters around \SI{3.1}{\nm} (\ac{TEM}, \autoref{fig:TEM_results}), they have very different hydrodynamic diameters (\ac{DLS}, \autoref{fig:DLS_results}), e.g. due to embedded cluster within a single dextran shell for Perimag. 
Clusters are known to be the \ac{MPI} active component for signal generation~\autocite{eberbeck_how_2011} of small \acp{MNP} like Perimag and our images show a similar core size of around \SI{5}{\nm} as reported by~\cite{eberbeck_multicore_2013}. Although the short-axis was counted in our work (see \autoref{sec:methods:TEM}) and the long-axis is more important for the measured magnetic properties, as it aligns with the easy axis, particle cluster and particle-particle interactions dominate the magnetic response in the \si{\kHz} range~\autocite{eberbeck_how_2011}. The effective \ac{MNP} size of Perimag is thus in a range of \SIrange{20}{50}{\nm} (visual inspection of \autoref{fig:TEM_results}) and these particle ensembles are then surrounded by the dextran shell which yields a hydrodynamic size at around \SI{100}{\nm} as reported in \autoref{fig:DLS_results} by \ac{DLS}.

Regarding Resovist and Resotran, \cite{gleich_fast_2010} reported that only \SI{3}{\percent} of the total iron mass are expected to be \ac{MPI} active, which agrees with our findings to the extent that the performance of Resotran/Resovist is better than Ferrotran, but worse than Perimag, suggesting that most particle ensembles are too small. Also, visual inspection of the \ac{TEM} images reveal small clusters in the \SIrange{10}{20}{\nm} range and Resotran/Resovist \acp{MNP} overlap much less than Perimag, with a hydrodynamic diameter of about \SI{65}{\nm}.
Accordingly, in the case of Ferrotran, the separation by the hull prevents any overlapping particles and particle-particle interactions are suppressed, resulting in \acp{MNP} that are too small for \ac{MPI} with an almost linear magnetization curve in the relevant excitation range. The magnetic properties of Ferrotran indicate that it is not eligible for \ac{MPI}, which was confirmed with images of poor quality, that do not resemble the measured phantoms in \autoref{fig:MPI_recos}.

The differences in saturation magnetization observed by \ac{VSM} most likely result from the different material composition ratios for the iron oxides magnetite and maghemite with a saturation magnetization of \SI{98}{\ampere\meter^2\per\kg_{Fe}} and \SI{82}{\ampere\meter^2\per\kg_{Fe}}, respectively~\autocite{colombo_biological_2012}. Additionally, the presence of clusters with particle-particle interaction contribute to a difference in the initial slope of the magnetization curve
as indicated by the \ac{VSM} data in \autoref{fig:VSM_results}.
Further, \ac{VSM} successfully predicted the magnetization curve that was measured using \ac{MPS}, without the hysteresis that is induced by the dynamic excitation field. The center plot of \autoref{fig:MPS_analysis} confirms this behavior, as the hysteresis curve widens with increasing excitation amplitude and the maximum saturation approaches the values measured with \ac{VSM} for high amplitudes, without surpassing them.
%%% pMPS 
The 1D \ac{MPS} results are supported by the achieved \ac{SNR} levels of the measured \ac{MPI} system matrices and indicate that the imaging quality of the tailored tracer Perimag outperforms all other tracers, which was eventually shown by the reconstruction of the spiral phantom.
%%% x-space
The analysis of \ac{MPS} data on tracer performance correlates well with the image reconstructions for all 4 tracers using the system matrix approach. We did not perform $x$-space reconstructions in the time-domain but expect our findings to generalize to other imaging sequences and reconstruction techniques. The \ac{FWHM} of the $x$-space \ac{PSF} also implies a good performance of Perimag and a bad performance of Ferrotran, indicated by a narrow and broad \ac{PSF}~\autocite{croft_relaxation_2012}, respectively.

Throughout all conducted measurements, the properties and performance of Resovist and Resotran proofed to be akin and the remaining differences could be due to variations between LOT numbers or due to different distributions of the iron oxides magnetite and maghemite.
The reconstructed images of \autoref{fig:MPI_recos} confirm suitability of Resotran for intricate \ac{MPI} applications and our work indicates its suitability for 2D sequences, where particle clustering is relevant. On this basis, suitability for more elaborate 3D sequences can be assumed, because the excitation direction is changing for a 2D sequence as well, which is the main difference of 2D/3D excitation to colinear 1D sequences. This is crucial for the medical translation of \ac{MPI}, although the performance of Resotran/Resovist does not reach the level of tailored MPI tracers. Cross-reconstructions of Resotran and Resovist are possible, emphasizing their resemblance, and surprisingly even improving the reconstruction result for Resovist using the Resotran system matrix. This might be due to a slightly higher \ac{SNR} of the Resotran SM, as indicated by the analysis in \autoref{fig:MPI_SM}. More investigations on this matter using different particle batches are necessary.

\section{Conclusion}

Resotran qualifies as a tracer for \ac{MPI} and its deployment in preclinical trials and system characterizations~\autocite{thieben_system_2023} could positively impact an official medical approval for \ac{MPI}. Furthermore, it will facilitate the process of clinical translation of \ac{MPI}, even if the signal performance is surpassed by tailored \acp{MNP}. Due to its similarity to the established \acp{MNP} Resovist in both, performance and composition, insights gained on Resovist are in principle transferable to Resotran.

%%%%%%%%%%%%%%%%%%%%%%%%%%%%%%%%%%%%%%%%%%%%%%%%%%%%%%%%%%%%%%%%%%%%%%%%%%%%%%%%%%%
\section*{Contributions~\&~Acknowledgments}

A.W. performed the DLS measurements and carried out the sample preparation for TEM.
The TEM measurements were performed by Stefan Werner at the Division of Electron Microscopy of the Chemistry Department at University of Hamburg under the direction of Dr. Charlotte Ruhmlieb.
J.A. performed the VSM measurements.
F.M. performed the MPS measurements and analysis.
K.S. measured the dilution series.
K.S. and F.M. performed the MPI measurements.
K.S. and T.K. reconstructed the MPI images and did the SNR analysis. 
T.K. supervised the project.
F.M., K.S., F.W., F.T., M.A., P.V., M.G. and T.K. contributed to the conceptualization and theory.
F.M., K.S., J.A., A.W. and F.W. wrote the original draft.
All authors reviewed the final manuscript.

%%%%%%%%%%%%%%%%%%%%%%%%%%%%%%%%%%%%%%%%%%%%%%%%%%%%%%%%%%%%%%%%%%%%%%%%%%%%%%%%%%%
\section*{Data Availability Statement}

The data that support the findings of this study are available upon reasonable request from the authors.

%%%%%%%%%%%%%%%%%%%%%%%%%%%%%%%%%%%%%%%%%%%%%%%%%%%%%%%%%%%%%%%%%%%%%%%%%%%%%%%%%%%
% \section*{Appendix}
% \label{sec:appendix}

%%%%%%%%%%%%%%%%%%%%%%%%%%%%%%%%%%%%%%%%%%%%%%%%%%%%%%%%%%%%%%%%%%%%%%%%%%%%%%%%%%%
\printbibliography%

\end{document}

%% file: acronyms.tex
\usepackage[nolist,nohyperlinks]{acronym}

\begin{acronym}[ECU]

\acro{AWMPS}[AWMPS]{arbitrary waveform magnetic particle spectrometer}
\acro{ADC}[ADC]{analog-to-digital converter}
\acro{AUC}[AUC]{area under the curve}

\acro{CT}[CT]{computed tomography}

\acro{DAC}[DAC]{digital-to-analog converter}
\acro{DFG}[DFG]{drive-field generator}
\acro{DFCs}[DFCs]{drive-field coils}
\acro{DF}[DF]{drive-field}
\acro{DA}[DA]{differential amplifier}
\acro{DLS}[DLS]{dynamic light scattering}
\acro{DTS}[DTS]{dispersion technology software}

\acro{ECD}[ECD]{equivalent circuit diagram}

\acro{FFL}[FFL]{field-free-line}
\acro{FFP}[FFP]{field-free-point}
\acro{FFR}[FFR]{field-free-region}
\acro{FFT}[FFT]{fast Fourier transform}
\acro{FOV}[FOV]{field-of-view}
\acro{FWHM}[FWHM]{full width at half maximum}

\acro{GUI}[GUI]{graphical user interface}

\acro{HCC}[HCC]{high current circuit}

\acro{ICN}[ICN]{inductive coupling network}
\acro{ISI}[ISI]{integrated signal intensity}
\acro{ICs}[ICs]{integrated circuits}
\acro{IA}[IA]{instrumentation amplifier}

\acro{ICU}[ICU]{intensive care unit} 
\acro{LFR}[LFR]{low-field-region}
\acro{LNA}[LNA]{low noise amplifier}

\acro{MPI}[MPI]{magnetic particle imaging}
\acro{MRI}[MRI]{magnetic resonance imaging}
\acro{MTT}[MTT]{mean-transit-time}
\acro{MNP}[MNP]{magnetic nanoparticle}
\acro{MPS}[MPS]{magnetic particle spectroscopy}

\acro{PSF}[PSF]{point spread function}
\acro{PNS}[PNS]{peripheral nerve stimulation}
\acro{PTT}[PTT]{pulmonary transit time}
\acro{PDI}[PDI]{polydispersity index}

\acro{Q}[Q-factor]{quality factor}

\acro{RF}[RF]{radio-frequency}
\acro{RP}[RP]{RedPitaya STEMlab 125-14}
\acro{RF-fields}[RF]{radio frequency fields}
\acro{rBV}[rBV]{relative blood-volume}
\acro{rBF}[rBF]{relative blood-flow}
\acro{rCBV}[rCBV]{relative cerebral-blood-volume}
\acro{rCBF}[rCBF]{relative cerebral-blood-flow}
\acro{RBCs}[RBCs]{red blood cells}

\acro{SNR}[SNR]{signal-to-noise ratio}
\acro{SU}[SU]{surveillance unit}
\acro{SAR}[SAR]{specific absorption rate}
\acro{SPIONs}[SPIONs]{superparamagnetic iron-oxide nanoparticles}
\acro{USPIONs}[USPIONs]{ultrasmall superparamagnetic iron-oxid nanoparticles}
\acro{SF}{selection field}
\acro{SFG}[SFG]{selection-field generator}
\acro{SEM}[SEM]{scanning electron microscope}
\acro{SSIM}[SSIM]{structural similarity index measure}

\acro{TF}[TF]{transfer function}
\acro{THD}[THD]{total harmonic distortion}
\acro{TTP}[TTP]{time-to-peak}
\acro{TEM}[TEM]{transmission electron microscopy}

\acro{USPIO}[USPIO]{ultra-small superparamagnetic iron oxide}

\acro{VOI}[VOI]{volume of interest}
\acro{VSM}[VSM]{vibrating sample magnetometry}

\end{acronym}